\documentclass[aps,prl,twocolumn,preprintnumbers,amsmath,amssymb,superscriptaddress]{revtex4-2}
\usepackage{graphicx}
\usepackage[caption=false]{subfig}
\usepackage{epsfig}
\usepackage{dcolumn}
\usepackage{bm}
\usepackage{color}
\usepackage{float}
\usepackage[colorlinks]{hyperref}
\usepackage{verbatim}
\usepackage{algorithm}
\usepackage{algorithmic}
\usepackage{tikz}
\usepackage{adjustbox}
\usetikzlibrary{quantikz}

\hypersetup{citecolor = blue}

\def\bra#1{\left\langle#1\right|}
\def\ket#1{\left|#1\right\rangle}

\def\Re{{\rm Re}}
\def\Im{{\rm Im}}

\def\be{\begin{equation}}       \def\ee{\end{equation}}
\def\bea{\begin{eqnarray}}      \def\eea{\end{eqnarray}}
\def\ba{\begin{array}}
	\def\ea{\end{array}}
\def\bnum{\begin{enumerate} }
	\def\enum{\end{enumerate}}

\def\=>{\Rightarrow}
\def\>{\rightarrow}

\def\eye2{Fathbb{I}}

\renewcommand{\v}[1]{{\bf #1}}

\renewcommand{\>}{\rangle}
\renewcommand{\Im}{{\rm Im}}
\renewcommand{\Re}{{\rm Re}}

\newcommand{\al}[1]{\begin{align}#1\end{align}}
\newcommand{\eq}[2]{
	\begin{equation}
		#1 \label{#2}
	\end{equation}
}

\renewcommand{\rm}[1]{\mathrm{#1}}

\usepackage{listings}
\usepackage{color}
\definecolor{lightgray}{gray}{1}

\begin{document}
	
\title{Universal imaginary-time critical dynamics on a quantum computer}
	
\author{Shi-Xin Zhang}
\email{shixinzhang@tencent.com}
\affiliation{Tencent Quantum Laboratory, Tencent, Shenzhen, Guangdong 518057, China}

\author{Shuai Yin}
\email{yinsh6@mail.sysu.edu.cn}
\affiliation{School of Physics, Sun Yat-Sen University, Guangzhou 510275, China}
	
	\begin{abstract}
 
      Quantum computers promise a highly efficient approach to investigate quantum phase transitions, which describe abrupt changes between different ground states of many-body systems. 
      At quantum critical points, the divergent correlation length and entanglement entropy render the ground state preparation difficult. In this work, we explore the imaginary-time evolution for probing the universal critical behavior as the universal information of the ground state can be extracted in the early-time relaxation process. 
      We propose a systematic and scalable scheme to probe the universal behaviors via imaginary-time critical dynamics on quantum computers and demonstrate the validness of our approach by both numerical simulation and quantum hardware experiments. With the full form of the universal scaling function in terms of imaginary time, system size, and circuit depth, we successfully probe the universality by scaling analysis of the critical dynamics at an early time and with shallower quantum circuit depth. 
      Equipped with quantum error mitigation, we also confirm the expected scaling behavior from experimental results on a superconducting quantum processor which stands as the first experimental demonstration on universal imaginary-time quantum critical dynamics.
      
	\end{abstract}

\date{\today}
\maketitle

{\it Introduction}---Quantum computing has been of great academic and industrial interest since Richard Feynman's original vision of using quantum systems to simulate nature. Recently, various experimental platforms of Noisy Intermediate-Scale Quantum (NISQ) \cite{Preskill2018}  devices have been developed, with the long-term vision for solving practical problems that a classical computer cannot address efficiently \cite{Arute2019, kim2023}.
Moreover, quantum algorithms for quantum simulation of ground state, excited state, and dynamical properties are expanding in recent years. Amongst these quantum simulation algorithms, variational quantum algorithms \cite{Bharti2021, Cerezo2020b, Endo2020, Tilly2021} are the most promising candidates in the NISQ era.
Various quantum error mitigation (QEM) techniques \cite{Cai2022, Temme2017, Czarnik2020, Lowe2021, Kim2021b, Strikis2021, Zhang2021d} have also been developed to alleviate the effect of quantum noise on NISQ devices and yield reliable experimental results. The rapid development of quantum hardware, quantum algorithms, and quantum software provides far-reaching platforms to investigate various exotic quantum phases.

As the watershed of different ground states, the critical points hold universal scaling behaviors, attracting extensive investigations as one of the cornerstones of modern physics \cite{Sachdev2011}. The routine method to explore the critical properties is to identify the ground state first and then calculate the physical quantities to reveal the scaling properties. However, for critical systems, difficulties are encountered in obtaining the ground state with quantum computers. The variational circuit ansatz should be carefully chosen to take into account the divergent entanglement as the divergence in general requires a divergent depth of quantum circuits to faithfully capture. 
In addition, the typical time scale to arrive at the ground state in imaginary-time evolution is proportional to $N^z$ in which $N$ is the size of the system and $z$ is the dynamic exponent. Therefore, for large systems, it takes an extremely long time to obtain the ground state accurately. 

In this work, we show that these disadvantages in probing quantum critical properties can be transformed into advantages by exploring universal scaling behaviors in the imaginary-time critical dynamics on quantum computers. Firstly, we identify that critical exponents appearing in the short-time dynamical scaling are the same as the static ones in the ground state. We can circumvent the need to get ground state based on this observation. Secondly, starting from a product initial state, both the correlation length and entanglement entropy are relatively small in the short-time stage. Therefore, we can infer the late-time (ground state) universal behavior via the short-time scaling at the early time. Finally, even the shallow variational circuits fail to faithfully reflect the imaginary-time dynamics, we can still extract the correct universal scaling based on finite-depth scaling. In sum, the universal properties of the critical point can be detected in an efficient and scalable way with short imaginary time evolved and shallow variational circuit depth required.

By using the powerful toolboxes for quantum simulation on NISQ devices, including the variational quantum eigensolver (VQE) \cite{Peruzzo2014, McArdle2020, Tilly2021, Zhang2021b, Liu2021c, Miao2023} , variational quantum dynamics simulation \cite{Li2017b, McArdle2019, Yuan2019, Lee2022}, and quantum error mitigation, we reveal the scaling form of imaginary-time critical dynamics on quantum computers in the one-dimensional quantum Ising model. Our results not only experimentally demonstrate the universal imaginary-time critical dynamics for the first time, but also pave the way for future studies on novel critical systems via imaginary-time relaxation dynamics on NISQ computers.


%
{\it Imaginary-time critical dynamics}---The imaginary-time evolution of a quantum system described by a Hamiltonian $H$ for a quantum state $|\psi(t)\rangle$ is given by the Schrödinger equation with an initial wave function $|\psi(0)\rangle$ as
$$
\frac{\partial}{\partial \tau}|\psi(\tau)\rangle=-H|\psi(\tau)\rangle,
\label{eqitse}
$$
imposed with the normalization condition $\langle\psi(\tau) \vert \psi(\tau)\rangle=1$. The formal solution can be regarded as a non-unitary evolution $e^{-\tau H}$ on the initial state:
$$
|\psi(\tau)\rangle=\frac{1}{Z} \exp (-H \tau)\left|\psi(0)\right\rangle,
$$
where $Z=\langle \psi(0)\vert \exp (-2H \tau)\left|\psi(0)\right\rangle$ is the normalization factor.

When $H$ is near its critical point, the universal scaling behaviors emerge in the imaginary-time relaxation process \cite{Janssen1989, Li1995st, Li1996st, Zheng1998, Yin2014, Zhang2014a, Shu2017}. From a product initial state with an initial order parameter $M_0$, the general scaling transformation of a physical quantity $P$ follows
\begin{equation}
   P\left(\tau, g, M_0, N\right)=b^\phi P\left[b^{-z}\tau, b^{1/\nu}g, U(b,M_0), b^{-1}N\right],
   \label{eqscalingtr}
\end{equation}
in which $b$ is the rescaling factor, $g$ is the distance in Hamiltonian parameter deviating from the critical point, $\phi$ is the scaling dimension of $P$, and $U(b,M_0)$ is a characteristic function \cite{Zheng1996}. For $M_0=1$, we have the fixed point $U(b,M_0)=1$. For very small $M_0$, $U(b,M_0)=M_0 b^{x_0}$. The critical exponents in Eq. (\ref{eqscalingtr}) should be equal to the equilibrium ones since they are connected via $\tau\rightarrow\infty$ limit, when Eq. (\ref{eqscalingtr}) should recover the equilibrium scaling for ground states.

By extending to the finite circuit depth case, where the static or dynamical states are prepared by a finite-depth variational quantum circuit, we have
\al{
	&P\left(\tau, g, M_0, N, D\right)\nonumber\\&=b^\phi P\left[b^{-z}\tau, b^{1/\nu}g, U(b,M_0), b^{-1}N,b^{-\alpha} D \right],
}
where $D$ is the circuit depth and $\alpha$ is its scaling dimension.

From a completely ordered initial state, for $P=M^k$, the $k$-th moment of the order parameter with $\phi=-k\beta/\nu$, by choosing $b=N$, one obtains the scaling form of $M^k$ at $g=0$ as,

\eq{M^k(\tau, N, D)  = N^{-k\beta/\nu}f_M (\tau N^{-z}).}{mscale}

When the finite depth effect is also considered, we have the scaling form at critical point $g=0$ as

\eq{M^k(\tau, N, D)  = N^{-k\beta/\nu}f_M (\tau N^{-z}, DN^{-\alpha}).}{eq:mv}

From a disordered initial state with the local order parameter distributed randomly and $M_0=0$, an imaginary-time correlator is defined as $A \equiv \overline{\left(\sum_{i=1}^N M^s_i(0)M^s_i(\tau)\right) }_s$, where the correlator is average over different initial spin configuration $s$ as well as qubit $i$. This correlator satisfies the dynamic scaling form as 
\eq{A=N^{-d+\theta z} f_A(\tau N^{-z}).}{}
in which $\theta$ is the critical initial slip exponent, unique to dynamical behavior, satisfying the scaling law
\eq{x_0 = \theta z + \beta /\nu.}{}

In this work, we use powerful NISQ toolboxes for quantum simulation including variational quantum eigensolver, variational quantum dynamics simulation, and quantum error mitigation to demonstrate the universal imaginary-time dynamics by identifying the scaling form and critical exponents discussed here. 

{\it Variational imaginary-time dynamics simulation}---There are two main proposals to enable the simulation of non-unitary imaginary-time dynamics on quantum computers. The ansatz free form, called QITE \cite{Motta2020}, implements the unitary approximation for each small Trotterized imaginary time step on the circuit progressively which requires an exponential large circuit depth. Therefore, this method doesn't scale well with system size and is not ready for NISQ devices even with only several qubits. 

The ansatz-based form, on the other hand, has a predefined variational circuit ansatz $U$. Given a set of variational circuit parameters $\vec{\theta}$, the output quantum state from the ansatz $\vert \phi(\vec{\theta}(\tau))\rangle=U(\vec{\theta})\vert 0 \rangle$ is taken as the variational quantum state under the imaginary time dynamics. For imaginary time $\tau$, by determining the optimal parameters $\vec{\theta}(\tau)$, we can obtain the quantum state $\vert \phi(\tau)\rangle$ and thus the relevant observables from the state. Therefore, the problem of simulating the dynamics of the quantum state is reduced to determining the dynamics of the circuit parameters $\vec{\theta}(\tau)$. 

There are two approaches to determine the circuit parameter dynamics. The first approach  \cite{McArdle2019, Yuan2019} utilizes the philosophy of  McLachlan’s variational principle \cite{McLachlan1964}.  The circuit parameters are evolved by an ordinary differential equation whose coefficients can be obtained given the knowledge of the variational quantum state $\vert \phi(\vec{\theta})\rangle$.  

The second approach of the ansatz-based family, called p-VQD \cite{Barison2021}, determines $\vec{\theta}(\tau)$ by constructing a variational optimization problem in each imaginary time step $d\tau$. In this approach, we tune $\vec{\theta}(\tau)$ to maximize the objective: $\langle \phi(\vec{\theta}(\tau)) \vert e^{-\tau H} \vert  \phi(\vec{\theta}(\tau-d\tau)) \rangle$. The small step of non-unitary evolution in between is easy to implement by embedding the non-unitary into a Hilbert space with an extra qubit and applying only one bit of post-selection \cite{Liu2020}. In the limit of $d\tau\rightarrow 0$, p-VQD recovers the result of McLachlan's variational principle, assuming the optimization problem is perfectly solved in each step. So the two approaches give identical circuit parameter dynamics trajectory in the ideal case. Numerically, p-VQD might be less stable than McLachlan's variational principle, as the former heavily relies on variational optimization with potential local minimum issue \cite{Bittel2021, Anschuetz2021, Liu2023a} while the latter directly gives the exact formula for the circuit parameter dynamics and avoid explicit optimization procedure. The optimization requires gradient descent where circuit parameter gradients are obtained via parameter shift scheme in experiments \cite{Li2017d, Schuld2019a, Banchi2020} or more efficiently simulated classically via automatic differentiation \cite{Bartholomew-Biggs2000a, GunesBaydin2018}. Therefore, we focus on  McLachlan's variational principle-based approach in this work. The caveat of ansatz-based implementation for imaginary-time dynamics is that the expressive power of the circuit ansatz can limit the approximation accuracy for the dynamics. However, we will utilize this aspect as finite circuit depth scaling which turns out to be helpful instead of harmful in identifying universal critical dynamics.

The time evolution dynamics for the circuit parameters under the Hamiltonian $H$ can be derived from McLachlan's variational principle \cite{Yuan2019}:
\begin{equation}
	\sum_j G_{i, j}^R \dot{\theta}_j=-C_i^R, \label{vmc}
\end{equation}
where $\theta$ is the parameters in the variational circuit and $\cdot^R$ is for taking the real part. 
The matrix of $G$ and the vector of $C$ are given by
\begin{equation}
	G_{i, j}=\frac{\partial\langle\phi(\vec{\theta}(\tau))|}{\partial \theta_i} \frac{\partial|\phi(\vec{\theta}(\tau))\rangle}{\partial \theta_j}, 
\end{equation}
\begin{equation}
	C_i=\frac{\partial\langle\phi(\vec{\theta}(\tau))|}{\partial \theta_i} H|\phi(\vec{\theta}(\tau))\rangle,
\end{equation}
respectively. The matrix elements can all be obtained from the real quantum hardware \cite{Yuan2019}. In the numerical simulation, the matrix $G$ and the vector $C$ can be much more efficiently obtained via unique features including vectorized parallel processing and automatic differentiation for Jacobians in TensorCircuit \cite{Zhang2022}.

We can solve the dynamics Eq. \ref{vmc} by regarding it as an ordinary differential equation problem with initial value $\theta_0$. We use the ODE solver with Runge-Kutta method provided by SciPy \cite{Virtanen2020} to solve the dynamics governed by Eq. \ref{vmc}. This ODE approach is more reliable and efficient compared to simple update given by discrete time steps $\epsilon$ as $\vec{\theta}(\tau+\epsilon) =\vec{\theta}(\tau) - \epsilon G^{-R} C^R $.

Throughout the work, we use the one-dimensional ($1$D) ferromagnetic coupled transverse field Ising model (TFIM) with open boundary conditions as the testbed, whose Hamiltonian is given as

\eq{H = \sum_{i=1}^N h X_i - \sum_{i=1}^{N-1} Z_iZ_{i+1}, }{}
And the quantum critical point is at $h=\pm 1$. The critical exponents for 1D TFIM are $\beta=1/8$, $\nu=1$, $z=1$ and $\theta\approx 0.373$ \cite{Yin2014}.

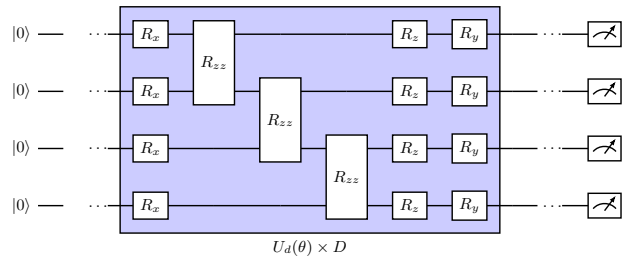
\begin{figure}[t]\centering
	\begin{adjustbox}{width=0.46\textwidth}
		\begin{quantikz}
			\lstick{$\ket{0}$} &\qw& \ldots &  \gate{R_{x}}\gategroup[4,steps = 6, style={fill=blue!20}, background, label style = {label position = below, anchor = north, yshift = -0.2cm}]{{\sc $U_d(\theta)\times D$}}     &\gate[2]{R_{zz}} &\qw &\qw &\gate{R_{z}} &\gate{R_{y}} &\qw & \ \ldots \qw& \meter{}\\
			\lstick{$\ket{0}$} &\qw & \ldots &\gate{R_{x}}  & & \gate[2]{R_{zz}}&\qw &\gate{R_{z}} &\gate{R_{y}}  &\qw& \ \ldots \qw& \meter{}\\
			\lstick{$\ket{0}$} &\qw & \ldots & \gate{R_{x}} & \qw & &\gate[2]{R_{zz}}&\gate{R_{z}} &\gate{R_{y}} &\qw & \ \ldots \qw& \meter{}\\
			\lstick{$\ket{0}$}  &\qw& \ldots  &\gate{R_{x}} & \qw& \qw& &\gate{R_{z}} &\gate{R_{y}} &\qw & \ \ldots \qw& \meter{}
		\end{quantikz}
	\end{adjustbox}
	\caption{The variational circuit ansatz for simulating imaginary time evolved state under 1D TFIM Hamiltonian. }
	\label{fig:ladder}
\end{figure}

The variational circuit ansatz we use is hardware efficient ansatz as follows

\eq{U(\theta) = \prod_{d=1}^D U_d(\theta_d),}{eq:ladder}
where for each circuit block we have (note that two-qubit gates are in ladder layout):

\eq{U_d(\theta_d) =\prod_{i=1}^N e^{-i\theta_{id4} Y_i}  e^{-i\theta_{id3} Z_i}\prod_{i=1}^{N-1}e^{-i\theta_{id2}Z_iZ_{i+1}}\prod_{i=1}^N e^{-i\theta_{id1} X_i}.}{ansatz1}
The circuit structure is schematic as Fig.~\ref{fig:ladder}.

{\it Numerical results}---At first we focus on the case for large enough $D$ such that the finite depth effects can be ignored. For the circuit ansatz and the system size $N$ of interest, we observe that the number of circuit blocks $D>N/2$ is in general sufficient to accurately represent the imaginary dynamics with relatively small errors (see the SM for details).

We first study the imaginary-time critical dynamics starting from $\ket{\uparrow^n}$ with $M_0=1$ in the imaginary time range $0\leq\tau\leq10$. By rescaling the order parameter calculated in variational circuit with $D\geq N/2$ as $M^2 N^{2\beta/\nu}$ and imaginary time $\tau$ as $\tau/N$, we find that the rescaled curves collapse well as shown in Fig. 2 (b), verifying Eq. (\ref{mscale}). The critical exponent is estimated as $\beta/\nu\approx 0.124\pm 0.03$, consistent with the exact value $\beta/\nu=0.125$. From Fig. 2, one finds that besides the long-time stage, scaling behaviors have already emerged at the short-time stage when the system is far away from the ground state, demonstrating that the critical properties can be detected in the short-time relaxation stage using quantum computers.

\begin{figure}[t]\centering
	\includegraphics[width=0.4\textwidth]{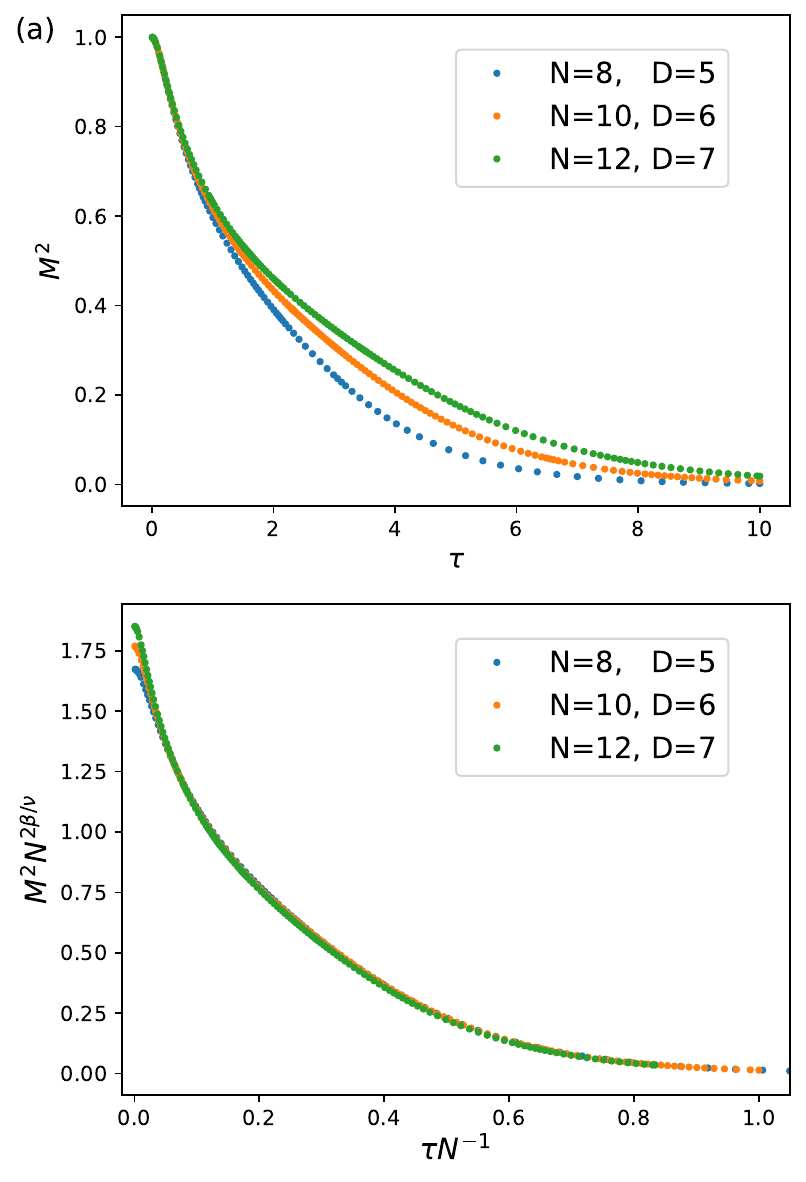}
	\caption{(a) Unscaled and (b) finite-size scaled data for variational imaginary-time dynamics using variational circuits of different size and depth $D=N/2+1$ starting from $M_0=1$ initial state. The scaling regime begins when $\tau > 0.1 N$. The critical exponent is estimated as $\beta/\nu\approx 0.124\pm 0.03$ is given $z=1$.}
	\label{fig:itescaling}
\end{figure}


Apart from the equilibrium critical exponent, the critical initial slip exponent $\theta$ can also be detected in the imaginary-time relaxation process in quantum circuits (see the SM for details).

\begin{figure}[t]\centering
		\includegraphics[width=0.43\textwidth]{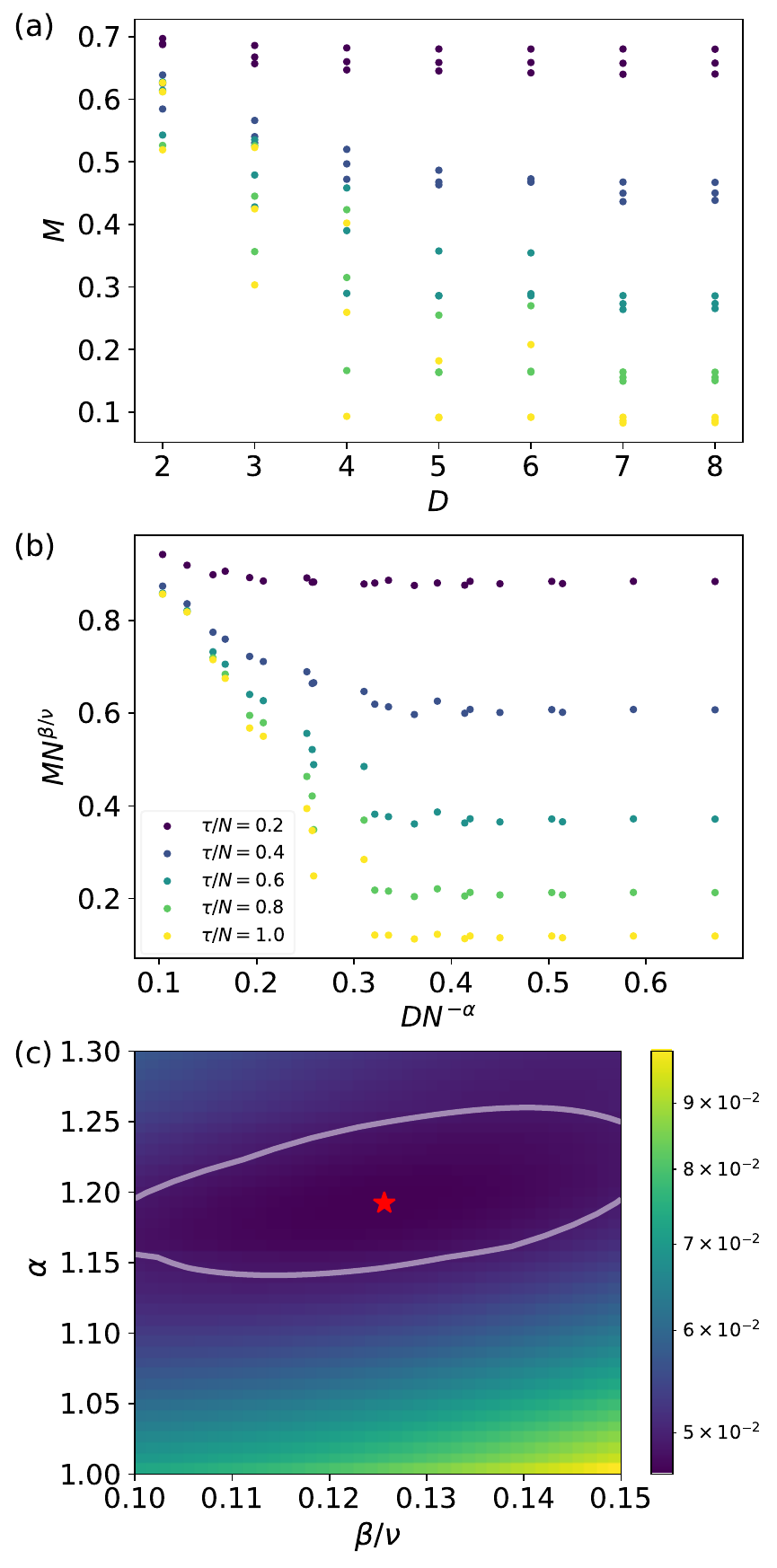}
		\caption{Imaginary-time dynamics data with several $\tau/N$ (a) unscaled and (b) rescaled according to the short-time finite-depth scaling. Points in the same color should fall into the same curve with the rescaled axis according to the finite depth exponent. (c) The data collapse quality for different guesses of finite-depth exponent $\alpha$ and critical exponent $\beta/\nu$. The best fit is estimated at $\alpha=1.19\pm 0.03$ and $\beta/\nu = 0.126\pm 0.04$. The grey contour indicates the region where the data collapse quality is no worse than $5\%$ compared to the optimal estimation.}
		\label{fig:twoscale}
\end{figure}

Then we explore the finite-depth effects in imaginary-time critical dynamics. It was shown that circuit depth $D$ also enters the general scaling form as finite-depth scaling \cite{Bravo-Prieto2020a, Jobst2022}
- a unique feature for simulation on quantum computers. In the imaginary-time relaxation process, we now verify Eq. \ref{eq:mv} directly and estimate the value of circuit depth exponent $\alpha$ as well as critical exponent $\beta/\nu$. We select several groups of data points with sharing $\tau N^{-1}$, i.e. fixing the first variable of Eq. \ref{eq:mv}. The data points are selected from circuits of depth $D$ from $2$ to $N/2+2$ with system size $N=8, 10, 12$. By rescaling $D$ and $M$ as $DN^{-\alpha}$ and $MN^{\beta/\nu z}$, we identify that for $\alpha\approx 1.19\pm 0.03$ and $\beta/\nu\approx 0.126\pm 0.04$, the rescaled curves collapse for fixed $\tau N^{-1}$, as shown in Fig.~\ref{fig:twoscale}, confirming the scaling behavior Eq. \ref{eq:mv}. We apply a systematic data collapse procedure to extract these exponents without any prior (see the SM for details). The extracted finite-depth exponent $\alpha$ is very consistent with the exponent reported in \cite{Jobst2022} for translational invariant infinite size circuits on the same model. Moreover, the value of $\alpha$ is further verified directly via the scaling form of the ground state by VQE simulation, where different circuit ansatzes give the same finite-depth exponent, demonstrating the universality of finite-depth scaling (see the SM for details).

From Fig.~\ref{fig:twoscale}, one finds that with fixed $\tau/N$, when $D=D_s \sim 0.3 N^\alpha$, the curves tend to saturate and are independent of $D$. Namely, for circuit depth larger than the saturated value $D_s\propto \tau^\alpha$, the approximation power for critical dynamics of the variational circuit is sufficient. More importantly, for circuits with shallower depth $D<D_s$, we can still extract these critical exponents from the data collapse procedure. In other words, due to finite-depth scaling, to extract critical exponents, the variational circuit required does not even need to faithfully capture the critical dynamics. We can obtain qualitatively correct estimations on these critical exponents from data at very early time and with very shallow circuits (see the SM for details).

\begin{figure}[t]\centering
	\includegraphics[width=0.5\textwidth]{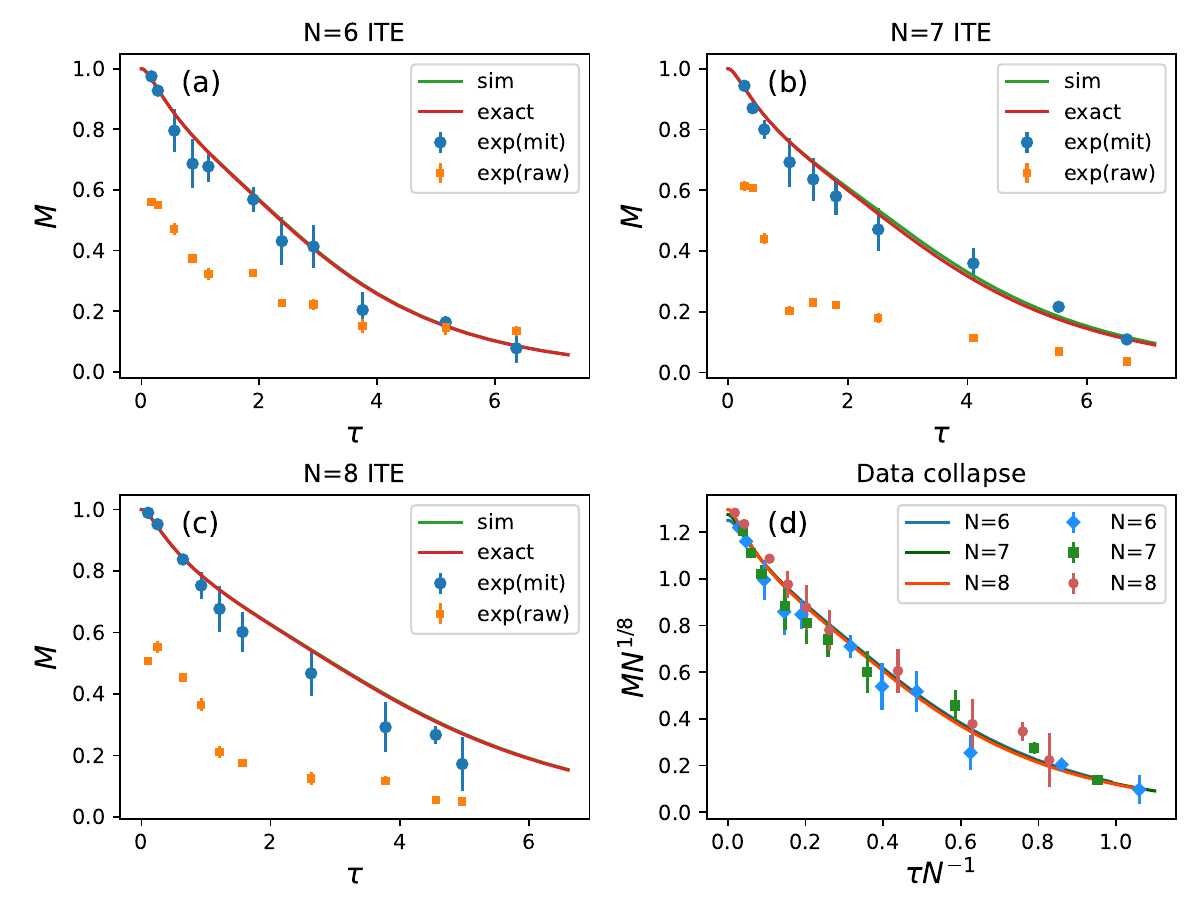}
	\caption{Experimental demonstration of the universal short-time quantum critical dynamics in imaginary time: the experimental data are collected from a programmable superconducting quantum processor. In (a), (b), (c) we show experimental results with only readout error mitigation (raw) and with extended Clifford data regression mitigation (mit) for $N=6,7,8$ qubits and $D=3,3,4$ variational circuit blocks, respectively. The exact results (exact) and numerical simulation results (sim) for the variational imaginary time evolution algorithm are also shown as lines for guidance. In (d), we apply data collapse on error mitigated data points to extract universal critical scaling from short-time dynamics, and the results coincide well with the theory prediction.}
	\label{fig:exp}
\end{figure}

{\it Quantum hardware experiments}---We also carried out experiments on a 20-qubit superconducting processor, and the variational imaginary-time dynamics results for $N=6,7,8$ as well as data collapse of them according to short-time critical dynamics are shown in Fig. \ref{fig:exp}.

We utilize a superconducting quantum processor with 20 transmon qubits to implement the critical dynamics. In experiments, we directly load the numerically exact circuit parameters for different time $\tau$ and evaluate the final magnetic order $M=\sum_{i}^N \langle Z_i\rangle$. 
Specifically, we firstly obtain the optimal circuit parameters $\vec{\theta}(\tau)$ for different imaginary time $\tau$ by numerical simulation in the noiseless limit and only evaluate the order parameter at given time $\tau$ with given circuit parameters $\vec{\theta}(\tau)$.

To accurately evaluate $M$ from quantum computers, several quantum error mitigation approaches \cite{Cai2022} are utilized in our experiment. We apply readout error mitigation on observable expectation level \cite{Bravyi2021b} and Clifford data regression (CDR)  \cite{Czarnik2020, Lowe2021} to mitigate both readout error and quantum gate error, respectively. The mitigated results with only the former method are labeled as \emph{raw} in Fig. \ref{fig:exp} while the mitigated results with both mitigation methods are labeled as \emph{mit} in Fig. \ref{fig:exp}. The accuracy for \emph{raw} results are not sufficient to observe universal imaginary-time dynamics due to the large quantum noise present on the quantum chip while the accuracy for \emph{mit} results can be helpful in identifying the finite-size short-time critical scaling for small system sizes as shown in \ref{fig:exp} (d). The result is the first experimental demonstration of short-time critical dynamics in the imaginary time direction. For a comparison of mitigated experiment results with unrescaled axis and finite-size rescaled axis, see the SM.

{\it Discussions}---Identifying quantum critical dynamics in the direction of imaginary time on quantum computers is not only of demonstration value but also paves a new way to investigate universal behaviors of critical quantum systems. Previously, to study the universal behavior of a critical system via variational quantum algorithms, one utilizes VQE to approximate the ground state at and near criticality to obtain the order parameter scaling behavior by measuring the observable from variational ground states. However, the ground state of a one-dimensional critical quantum system has a logarithmic law entanglement entropy $S\sim \ln N$, which requires a great circuit depth to fully capture. Instead, via universal quantum imaginary-time dynamics, the scaling behavior with the same sets of critical exponents can be revealed at very early time $\tau \ll N$ with shallow circuits having larger approximation errors. At that early stage, the half-chain entanglement scaling is given by the universal behavior $S\sim \ln\tau$ starting from zero, much less than the ground state case. In addition, the depth of the variational circuit can be further reduced as reaching entanglement of $S\sim \ln \tau$ is also not necessary thanks to the finite-depth scaling. Therefore, via the lens of universal imaginary-time critical dynamics, much less quantum computational resources are sufficient to investigate critical phenomena compared to ground state simulation. Note that these advantages of imaginary-time critical dynamics can also manifest themselves in higher dimensional critical systems. 

It is also an interesting future direction to study the scaling behavior of variational imaginary time dynamics with intrinsically distinct circuit architectures such as multi-scale entanglement renormalization ansatz \cite{Vidal2008, Evenbly2009, Kim2017a, Sewell2023} or with dynamically changed circuit structures via adaptive scheme \cite{Yao2020} or architecture search scheme \cite{Zhang2020b, Zhang2021, Lu2020}.

In sum, we propose a new scalable approach to study quantum critical behavior on quantum computers and systematically investigate the universal imaginary-time dynamics on quantum computers with extensive numerical simulation and quantum hardware experiments.

\section{Methods}

{\bf Quantum Software Framework:} All the high performance numerical simulation, as well as the quantum hardware experiments in this work are conducted with TensorCircuit \cite{Zhang2022}: an open-source, high-performance, full-featured quantum software framework for the NISQ era. The long-term vision of TensorCircuit is to unify the infrastructures and paradigms of quantum programming by providing unified backends, unified devices, unified providers, unified resources, unified interfaces, unified engines, unified representations, and unified pipelines. The software can simulate the quantum circuit with advanced tensor network contraction engine and supports modern machine learning engineering paradigms: automatic differentiation, vectorized parallelism, just-in-time compilation and GPU compatibility. The software also supports CPU/GPU/QPU hybrid deployment with an integrated quantum error mitigation toolbox for quantum hardware SDK.

{\bf Quantum Error Mitigation:} To evaluate $M$, the sum of expectation of local Pauli Z operators, from quantum computers, we utilize two methods to mitigate the errors. We firstly apply scalable readout error mitigation on observable expectation level natively assuming local tensor product structure of the readout error \cite{Bravyi2021b}. This approach works well in practice since the readout error on the device is well approximated by local structures with very little readout error correlation across qubits. We label the experimental results with only readout error mitigation {\emph {raw}}. To apply such readout error mitigation on expectations, we use the built-in readout error mitigator for observable expectations in TensorCircuit \cite{Zhang2022}. The exact formula for the readout error mitigation in this case can be derived analytically. Suppose the target observable is Pauli Z operator on qubit $i$ as $\langle Z_i \rangle$ and the readout error rates for 0 to 1 and 1 to 0 are $\epsilon_i$ and $\eta_i$ on qubit $i$, respectively. Note that these local readout error rates can be calibrated via simply running two benchmark circuits. For each readout result of 0 state on qubit $i$, we count the contribution to $\langle Z_i\rangle$ as $\frac{1+\epsilon_i-\eta_i}{1-\epsilon_i-\eta_i}$  instead of simply $+1$. Similarly, for each readout result  of 1 state on qubit $i$, we count the contribution to $\langle Z_i \rangle$ as $-\frac{1-\epsilon_i+\eta_i}{1-\epsilon_i-\eta_i}$ instead of $-1$. Such a formula is the direct consequence of Eq. (6) in Ref. \cite{Bravyi2021b}.

The accuracy for {\emph {raw}} results with only readout error mitigation is not sufficient to observe universal imaginary-time dynamics due to the quantum noise on the quantum chip.  We further apply Clifford data regression (CDR) approach  \cite{Czarnik2020, Lowe2021} to mitigate quantum errors and obtain reliable expectation estimation for order parameter $M$. The basic idea behind CDR is to firstly build several similar near Clifford circuits close to the target circuit to be evaluated. We then run each near Clifford circuit on both quantum hardware and the classical simulator to obtain two sets of results $M_{\text{noisy}}$ and $M_{\text{ideal}}$ for each near Clifford circuit instance. Via the data of  $M_{\text{noisy}}$ and $M_{\text{ideal}}$, we can fit a linear regression relation by the least square method, i.e. $M_{\text{ideal}} = a M_{\text{noisy}} +b$ where $a, b$ are learning parameters. Finally, by running our target circuit on the quantum hardware with the results as $M_{\text{noisy}}$, we can recover $M_{\text{ideal}}$ for the target circuit via the linear regression relation. In our experiment, we build several groups of circuit samples with different ratios of non-Clifford gates, and train them together to obtain the linear relation. We call this specific method extended CDR. We believe training on data with multiple non-Clifford ratios can make CDR results more robust and reliable. For each group $i$, we build $n_{i}$ circuits by uniformly replacing approximately $1-r_i$ ratio of non-Clifford single-qubit gates to the closest Clifford gate (in terms of Rz rotation angles). For small time scale $\tau< 1$, the prediction inaccuracy on the hardware is relatively small, so we use $n_1=5, r_1=0.6$, $n_2=5, r_2=0.7$, $n_3=5, r_3=0.8$ and $n_4=5, r_4=0.9$, four groups and $20$ circuits in total to learn the linear relation between the noisy prediction of $M$ on the chip and the ideal expectation of $M$ simulated classically. For larger $\tau$, the true value of $M_{\text{ideal}}$ is smaller and the experiment accuracy becomes worse, so we use $n_1=5, r_1=0.8$, $n_2=5, r_2=0.9$, $n_3=5, r_3=0.95$ three groups and $15$ circuits to learn the linear prior. The results obtained using this CDR pipeline are labeled as {\emph {mit}}. To apply CDR error mitigation technique, we use the CDR method from Mitiq \cite{LaRose2022} with further customization on multiple non-Clifford ratio support and TensorCircuit compatibility. It is worth noting that the specific CDR approach we adopt here has a very high ratio of $r_i$ on average indicating scalability issues. A high ratio of non-Clifford gate can lead to a circuit data with similar $M$ as the original circuit which greatly improves the accuracy for {\emph {mit}} results in our experiment. In other words, a very low ratio of non-Clifford gates such as $r=0.1$ is not sufficient to give stable and accurate mitigated experimental results now due to the relatively large quantum noise, especially cross-talk error present on the current generation quantum chip. Such high ratios cannot maintain for larger system sizes when stabilizer circuit simulation formalism is required where the simulation complexity is exponential with the number of non-Clifford T gates. Our perspective is, with further development of the quantum hardware, the experiment will be impacted by less quantum noise and the corresponding non-Clifford ratio $r$ in CDR can go down to the classical simulatable regime even with larger system size in the future. 

	~\newline
	\textbf{Acknowledgements:}
	We would like to acknowledge the helpful discussion with Hong Yao. S.-.X. Zhang would like to acknowledge the discussion and collaboration on a related error mitigation project with Yu-Qin Chen.  S. Yin is supported by the National Natural Science Foundation of China, Grants No. 12222515 and No. 12075324.
%

	\clearpage
	
	\begin{widetext}
		\section*{Supplemental Materials}
		\renewcommand{\theequation}{S\arabic{equation}}
		\setcounter{equation}{0}
		\renewcommand{\thefigure}{S\arabic{figure}}
		\setcounter{figure}{0}

\subsection{Initial parameters in the variational dynamics simulation}	

In this section, we show that with the circuit ansatz in the main text and TFIM Hamiltonian, the variational dynamics simulation is constrained and fails to reproduce the correct quantum imaginary-time dynamics if the initial circuit parameters are all strictly zero at the beginning (starting from perfect $\vert \uparrow^N\rangle$ state). 

Recall the time evolution dynamics for the circuit parameters under the Hamiltonian $H$ based on McLachlan's variational principle:
\begin{equation}
	\sum_j G_{i, j}^R \dot{\theta}_j=-C_i^R, \label{}
\end{equation}
where $\theta$ is the parameters in the variational circuit. 
The matrix of $G$ and the vector of $C$ are given by
\begin{equation}
	G_{i, j}=\frac{\partial\langle\phi(\vec{\theta}(\tau))|}{\partial \theta_i} \frac{\partial|\phi(\vec{\theta}(\tau))\rangle}{\partial \theta_j}, 
\end{equation}
\begin{equation}
	C_i=\frac{\partial\langle\phi(\vec{\theta}(\tau))|}{\partial \theta_i} H|\phi(\vec{\theta}(\tau))\rangle,
\end{equation}
respectively. 

We consider the general case when only circuit parameters of Ry gates are nonzero. If we can show that in this case $C$ are zero everywhere except at the Ry gates position, we know that the circuit parameter can only evolve nontrivially on Ry parameter subspace which fails to capture the imaginary-time dynamics variationally. This failure is not due to the low expressive power of the ansatz as the ansatz can correctly express the evolved state given appropriate circuit parameters. 
Instead, the failure is from the interplay between the initial parameters and the Hamiltonian which we call the phenomena variational dynamical constraint. Since we study the system evolved from $\vert 0^N\rangle$ ($M=1$) initial state, all initial circuit parameters are zero in the given ansatz and fall into the category of variational dynamical constraint failure. Therefore, to correctly characterize the dynamics in the numerical simulation, we perturb the initial circuit parameter from zeros at the beginning, where the perturbation is small enough to not affect the correctness of the dynamics and large enough to avoid the variational dynamical constraint failure.

Suppose we evaluate the $i$-th element of $C$ vector and the $i$-th parameter is binding to a Pauli operator $P=\prod_{i\in S}P_i$ as a rotation gate $e^{-i\theta_i P}$ which is not $Y$ ($C$ elements corresponding to Ry gate can have nonzero amplitude), we have
\al{C_i &= \Re{\langle 0 \vert  \prod_{i=0}^{N-1}e^{i(\theta_i+\theta'_i) Y_i} H   \prod_{i=0}^{N-1}e^{-i\theta'_i Y_i}\cdots (-iP) \prod_{i=0}^{N-1}e^{-i\theta_i Y_i}\vert 0\rangle}\nonumber\\
&\propto \Im \langle \phi \vert H \prod_{i\in S} (P_ie^{-2i\theta_i'Y_i})\vert \phi\rangle,  \label{eq:impart}}
where we have $\vert \phi\rangle =e^{-i(\theta_i+\theta'_{i}) Y_i}\vert 0\rangle $, $\langle \phi \vert Y_i\vert \phi\rangle =0$. We can easily check that $C_i$ in Eq. \ref{eq:impart} is zero for $P=X_i$, $P=Z_i$ and $P=Z_{i}Z_{i+1}$ which correspond gate types in the ansatz we use. Therefore, the dynamics can only evolve in the subspace of Ry gate freedom if the initial condition is strictly zero for all other gate parameters. And from the derivation, we clearly see that the simulation failure is not from symmetry argument, instead it is determined by the special interplay of several factors: the form of the ansatz, the form of the Hamiltonian, and the initial parameter choice. 

\subsection{Variational dynamics error with different circuit depth $D$}

Fig.~\ref{fig:iteerror} shows the variational dynamics simulation error in terms of $\delta M^2$ compared to the analytic exact imaginary-time dynamics results obtained by exact diagonalization. The system Hamiltonian is 10-qubit 1D TFIM with open boundary conditions. We conclude that the approximation is good enough for the circuit depth $D>N/2$ at least in the system size range that we explored in this work.

\begin{figure}[h]\centering
	\includegraphics[width=0.5\textwidth]{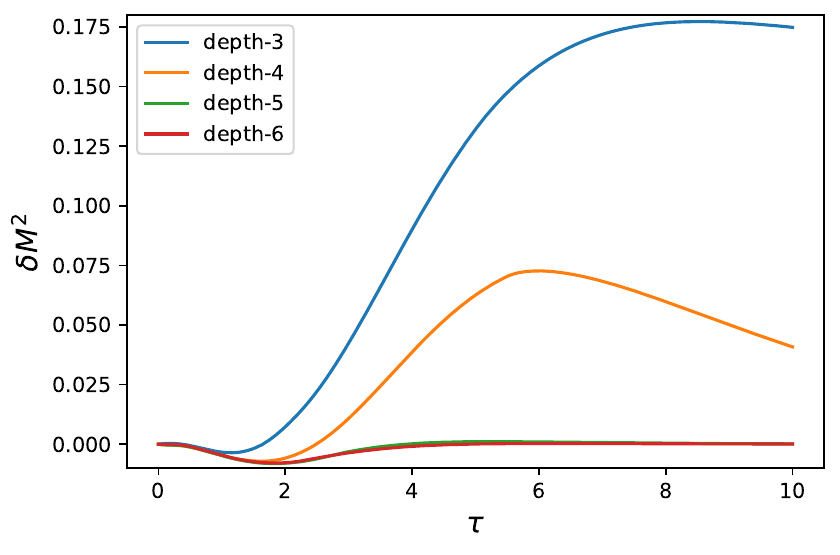}
	\caption{The absolute error in terms of $M^2$ for $N=10$ system variational quantum dynamics simulation of different circuit ansatz depth quenched from $\ket{\uparrow^n}$. The error is relatively small and saturate when the circuit depth exceeds half of the system size $N$.}
	\label{fig:iteerror}
\end{figure}

\subsection{Finite-depth scaling extracted from ground state simulation}

To disentangle different scaling factors (imaginary time $\tau$ and circuit depth $D$) and better understand them, we directly utilize variational ground state simulation to extract the finite-depth critical exponent. The ground state can be taken as the limit of infinite time $\tau\rightarrow \infty$ in the context of imaginary time evolution and we expect the scaling form as 
\eq{M^2(L, D)= N^{-2\beta/\nu} f_M(N^{-\alpha}D).}{eq:depth}

To solve the ground state problem, the variational quantum eigensolver (VQE) algorithm is utilized. In VQE, we directly minimize $E(\theta) = \bra{0}U(\vec{\theta})^\dagger H U(\vec{\theta})\ket{0}$ by tuning the parameters $\theta$ based on gradient descent, i.e. $\theta = \theta -\epsilon \frac{\partial E(\theta)}{\partial \theta}$, where $\epsilon$ is the learning rate. These circuit parameter gradients can be obtained via the parameter shift scheme in experiments and be more efficiently simulated classically via automatic differentiation integrated with TensorCircuit.

We use the same Hamiltonian model and circuit ansatz as given in the main text to run the VQE. For each system size $N$ and circuit depth $D$, we run independent optimization over $64$ different random initialization parameters sampled from Gaussian distribution with center $0$ and standard deviation $0.1$. The optimal final results of these $64$ trials are reported to avoid local minimum issues. We run the gradient descent 10000 steps to ensure the convergence with Adam optimizer and an exponential decay learning rate schedule. The hyperparameter of the optimizer is tuned for better convergence speed and accuracy. The order parameter $M^2$ of the converged state with different sizes $N$ and depth $D$ are shown in Fig.~\ref{fig:gsv1}(a). And we can do a finite-size finite-depth scaling on the data to extract critical exponent for depth $\alpha$ according to Eq. \eqref{eq:depth}, see Fig.~\ref{fig:gsv1}(b). The exponent is estimated as $\alpha=1.28\pm 0.06$. This value is very similar to the exponent reported in \cite{Jobst2022} where $1.21, 1.07$ are estimated for translational invariant infinite size circuit on 1D TFIM model for entanglement entropy and order parameter, respectively. To demonstrate such finite depth exponent is universal, we apply VQE with very different circuit ansatz on the same Hamiltonian, and the result on $\alpha$ is consistent as explained below. The $\alpha$ estimated from VQE here is also very close to the result in the main text extracted from the dynamics simulation.


\begin{figure}[t]\centering
	\includegraphics[width=0.38\textwidth]{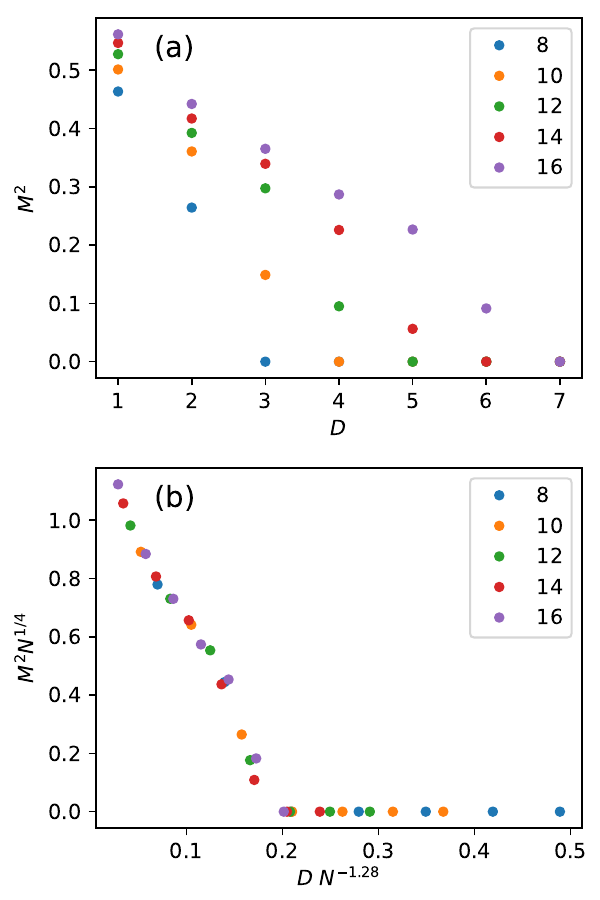}
	\caption{(a): $M^2$ of converged quantum state after VQE optimization on quantum computers of different sizes $N$ and depths $D$. (b): Data collapse for the converged state on quantum computers of different sizes and depth. We estimate the critical exponent for the depth scaling as $\alpha=1.28\pm 0.06$.   }
	\label{fig:gsv1}
\end{figure}

To demonstrate the universality of the finite-depth exponent $\alpha$, we also carry out VQE with different circuit ansatz. The alternative ansatz has brickwall two-qubit layout and fewer density of single-qubit gates and hence less expressive power with the same depth $D$. Specifically, the alternative circuit structure (see Fig. \ref{fig:brick}) is composed of $D$ blocks of Rx gates and two-qubit gates $e^{i \theta Z_iZ_{i+1}}$ are placed in an even-odd brickwall fashion instead of the ladder layout in the main text.
Each block $U_d$ can be expressed as:

\eq{U_d(\theta_d)=\prod_{i\in \text{odd}}^N e^{-i \theta_{id2}Z_{i}Z_{i+1}}\prod_{i\in \text{even}}^N e^{-i \theta_{id2}Z_{i}Z_{i+1}} \prod_{i=1}^N e^{-i \theta_{id1} X_i}.}{}

The extracted depth critical exponent $\alpha$ in this case is consistent with the former ansatz, implying a universal finite depth exponent for VQE of TFIM Hamiltonian. The result is summarized as Fig. \ref{fig:gsv3}.  

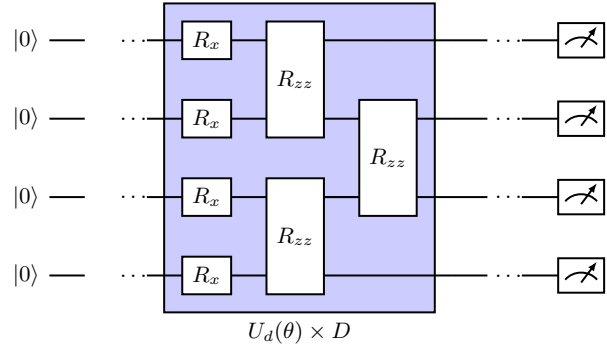
\begin{figure}[t]\centering
	\begin{adjustbox}{width=0.45\textwidth}
		\begin{quantikz}
			\lstick{$\ket{0}$} &\qw& \ldots &  \gate{R_{x}}\gategroup[4,steps = 3, style={fill=blue!20}, background, label style = {label position = below, anchor = north, yshift = -0.2cm}]{{\sc $U_d(\theta)\times D$}}     &\gate[2]{R_{zz}} &\qw  &\qw & \ \ldots \qw& \meter{}\\
			\lstick{$\ket{0}$} &\qw & \ldots &\gate{R_{x}}  & & \gate[2]{R_{zz}}&\qw   & \ \ldots \qw& \meter{}\\
			\lstick{$\ket{0}$} &\qw & \ldots & \gate{R_{x}} & \gate[2]{R_{zz}}& \qw &\qw & \ \ldots \qw& \meter{}\\
			\lstick{$\ket{0}$}  &\qw& \ldots  &\gate{R_{x}} & & \qw&\qw   & \ \ldots \qw& \meter{}
		\end{quantikz}
	\end{adjustbox}
	\caption{The alternative variational circuit ansatz for VQE to demonstrate the universal finite-depth scaling. Each block is composed of only one layer of Rx gates and one layer of Rzz gates in brickwall layout. }
	\label{fig:brick}
\end{figure}

\begin{figure}[t]\centering
	\includegraphics[width=0.38\textwidth]{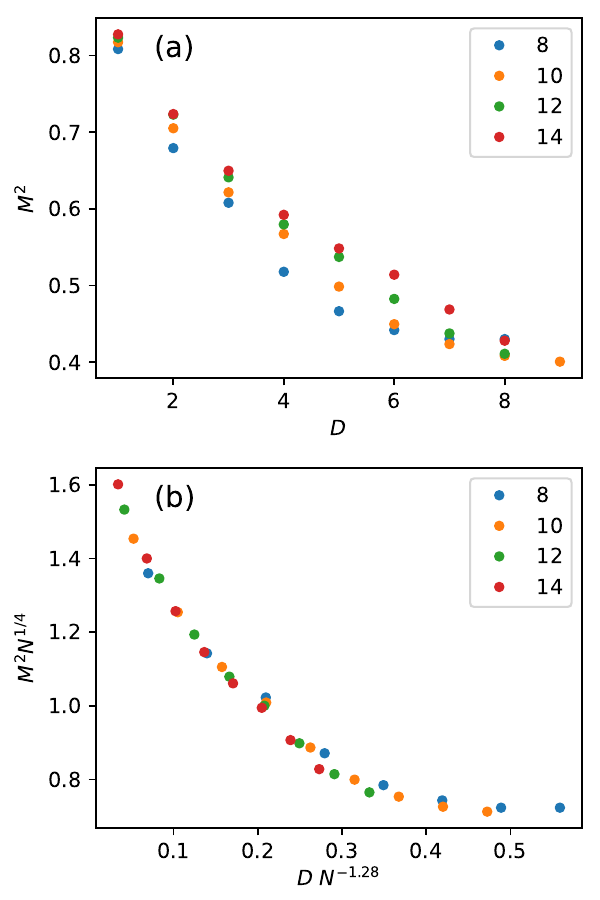}
	\caption{(a): $M^2$ of converged quantum state after VQE optimization on quantum computers of different sizes $N$ and depth $D$ with alternative ansatz. (b): Data collapse  for the converged state on quantum computers of different sizes and depths. The critical exponent for the depth scaling is $\alpha=1.28\pm 0.07$, consistent with the circuit ansatz in the main text.   }
	\label{fig:gsv3}
\end{figure}

\subsection{Hyperparameters in numerical simulation}

For variational quantum dynamics simulation, all zero initial parameters are perturbed with a Gaussian distribution with center zero and standard deviation $0.002$ to avoid the constraint on Hilbert space of only Ry rotation as discussed before.
The quantum Fisher information  matrix in the parameter dynamics equation can have very bad condition number, so we add $10^{-7} I$ to the matrix before getting its inverse to increase the numerical stability. We use RK45 ODE solver in scipy with default settings and a relative tolerance $10^{-4}$.

For variational quantum ground state simulation, we run the gradient descent 10000 steps to ensure the convergence with Adam optimizer. We design a learning rate schedule that exponentially decays from 0.02 with a decay step 2000 for a $60\%$ drop. Namely the learning rate at iteration step $i$ is controlled by $\epsilon(i) = 0.02 * 0.6 ^{i/2000} $. The hyperparameter of the optimizer is tuned as such for better convergence speed and accuracy. We run 64 sets of different random initialization from a Gaussian distribution center at zero with a standard deviation $0.1$, among these converged results, the best one is reported as the final converged value to avoid local minimum.

\subsection{Comparison of experiment results before and after data collapse}

The mitigated results from the quantum hardware experiment are presented before and after the correct data collapse (axis finite-size rescaling). See Fig. \ref{fig:exp2cols}.

\begin{figure}[t]\centering
	\includegraphics[width=0.7\textwidth]{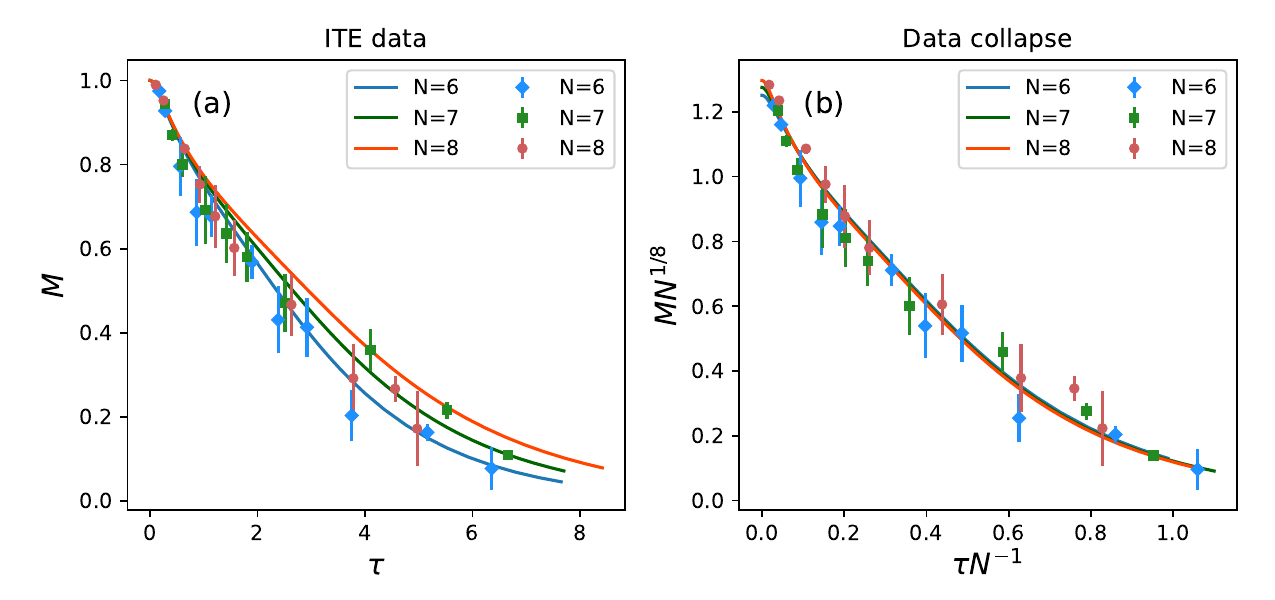}
	\caption{(a) Experimental results from quantum computers via error mitigation. (b) The data collapse. All data points are expected to fall into the same curve in the ideal case.}
	\label{fig:exp2cols}
\end{figure}
	\subsection{Experimental Hardware}
All experiments were performed on the 20-qubit 
quantum device. 
The topology of the device is a $10\times2$ grid, see Fig. \ref{fig:20xmon}. In the experiment, we only utilize the second row of the qubits (Qubit 12-19 for 8-qubit experiment, qubit 13-19 for 7 qubit-experiment, and qubit 14-19 for 6 qubit-experiment). Typical mean error rates of qubit 12-19
are $1.6 \times 10^{-2}$ for two-qubit gates, $0.14 \times
10^{-2}$ for single-qubit gates and $7\times 10^{-2}$ for
readout errors. Mean $T_1$ and $T_2$ time for qubit 12-19 are $24\mu s$ and $5\mu s$, respectively. 

\begin{figure}[t]\centering
	\includegraphics[width=0.6\textwidth]{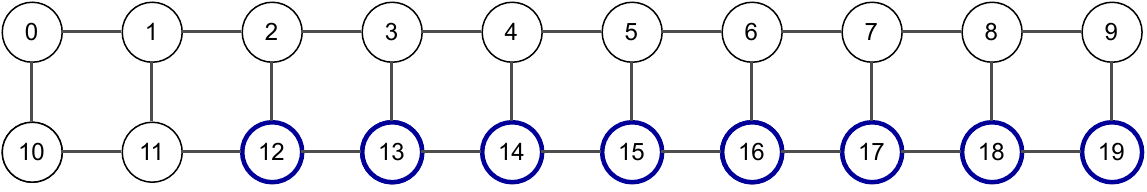}
	\caption{The qubit layout and coupling map for the 20-qubit device. The qubits with blue circles are used in the experiment of this work.  }
	\label{fig:20xmon}
\end{figure}

\subsection{Scaling results of imaginary-time correlator}

We study imaginary-time-correlator $A$ from a set of disordered initial state $M_0=0$ via the variational quantum imaginary-time dynamics simulation for system size from $N=8, 10, 12$ and circuit depth $D=N/2+1$ which is sufficient to capture the exact imaginary-time dynamics. The initial state is determined by randomly flipping (applying X gates) on half of the qubits at the beginning of the circuit. The results are consistent with the critical exponent $\theta=0.373$ for 1D TFIM from the data collapse according to Eq. (5) as shown in Fig.~\ref{fig:sm_a}.

\begin{figure}[t]\centering
	\includegraphics[width=0.45\textwidth]{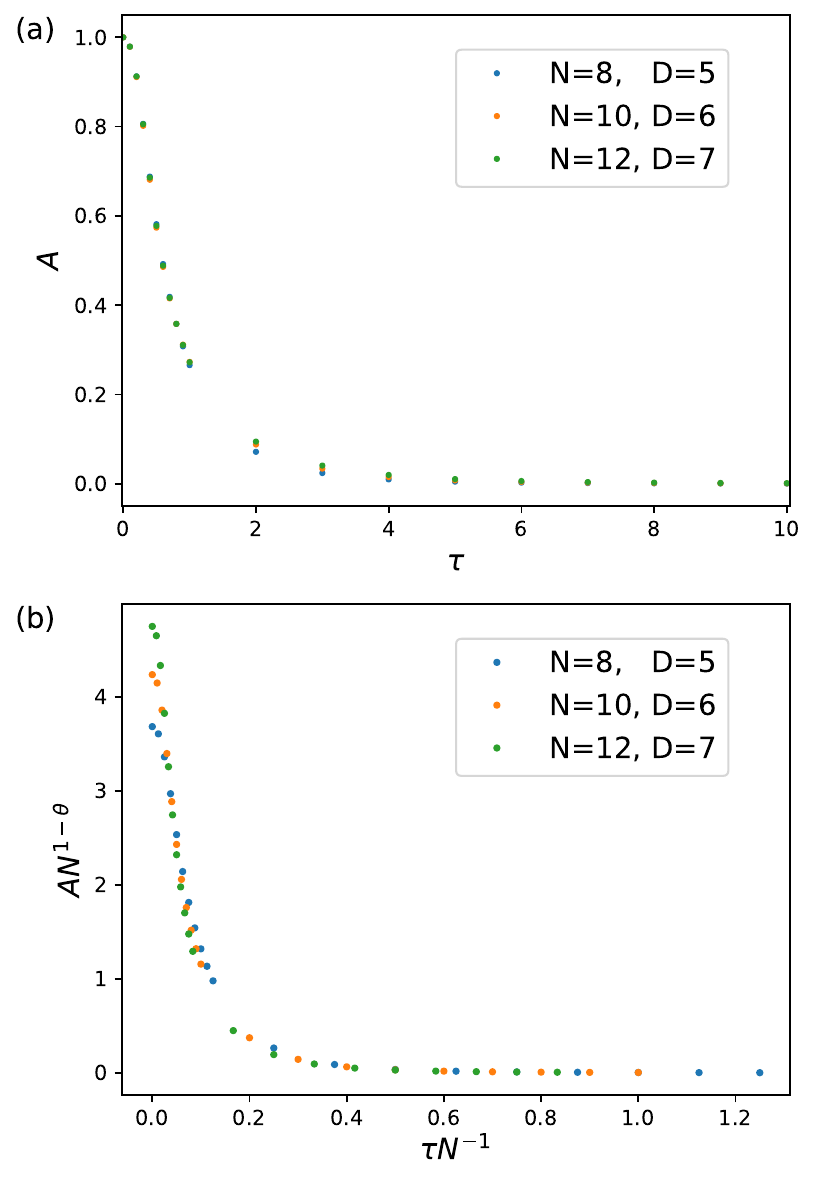}
	\caption{Imaginary-time correlator $A$ (a) unscaled and (b) finite-size scaled. The imaginary-time dynamics related critical exponent $\theta\approx 0.373$.}
\label{fig:sm_a}
\end{figure}

\subsection{Finite size scaling analysis}

The scaling analysis procedure is similar to the approach presented in Appendix A in \cite{Skinner2019a}. We use the fit on data in Fig. 3 in the main text as an example to showcase the workflow, the fit on data in other figures is similar. 

For a set of estimated exponents of $\beta/\nu$ and $\alpha$, one can define a cost function $R\left(\beta/\nu, \alpha\right)$ and try to minimize the cost by searching optimal values for $\beta/\nu$ and $\alpha$.  First, we rescale the data in each group $(D_i, M_i)$ (fixed $\tau/N$, different size $N$) as $x_i = DN^{-\alpha}$ and $y_i=MN^{\beta/\nu}$, leading to a family of curves $y_N(x)$, one curve for each system size $N$. The cost function $R$ is defined as the sum of the mean-squared deviations of each curve from their mean, summed over all unique points $x_i$ in the data set. In other words,
$$
R=\sum_{i, N}\left[y_N\left(x_i\right)-\bar{y}\left(x_i\right)\right]^2,
$$
where $y_N\left(x_i\right)$ indicates the value of $y_N$ at the point $x_i$. If this value is not specified explicitly in the data, it can be estimated by linear interpolation. Note that $\bar{y}(x)$ is the mean value of $y_N(x)$ over different system sizes $N$. For multiple variable scaling fit such as Fig. 3 in the main text, we also need to sum the cost function for each group of different $\tau/N$. The physical meaning is that we expect to identify the suitable exponents such that data points of the same $\tau/N$ lie on the same curve while points of different $\tau/N$ belong to different curves.

Given the numerical data, we can locate the best guess on critical exponents by extensive grid search since there are only two variables. The grid search result is similar as Fig. 3(c) in the main text and we can extract the optimal estimation for the exponents as well as the error bar by specifying some threshold on the cost function.

Specifically, for the data collapse of Fig. 2 in the main text, we use imaginary time evolution data from $N=8,10,12$ and circuit depth $D=N/2+1$ at $0.2\leq \tau\leq 10$. We assume a prior of $z=1$ and fit the single exponent $\beta/\nu$ by grid search.

For the data collapse of Fig. 3, we also assume $z=1$ as a prior, and try to identify the best $\beta/\nu$ and finite depth exponent $\alpha$ at the same time. We use the numerical results from seven curves $\tau/N=0.1,0.2,0.3,0.4,0.5,0.6, 0.7$, with system size $N=8,10,12$ and circuit depth $2\leq D\leq 8$ to fit the scaling behavior. 

\subsection{Scaling analysis with data from an even earlier time and shallower circuit}

For the scaling analysis on data from Fig. 3, we note that even if we only include data with short time and small depth, we can still capture the qualitatively correct scaling behavior. For example, if we only include data with $\tau/N=0.1, 0.2, 0.3, 0.4$ and circuit depth $2\leq D\leq 5$, we can still obtain similar critical exponent estimation as shown in Fig. \ref{fig:sm-small-scaling}. The estimated result deviates from the exact value a little. But considering how shallow the circuit is and how short time we are using, the relatively reasonable estimation we obtain demonstrates the key point of this work: we can access the universal critical properties via finite-size, short-time and finite-depth scaling. Via finite-size scaling, we can probe the thermodynamic behavior via small size system; via short-time scaling, we can probe the ground state behavior at the early imaginary time stage; and via finite-depth scaling, we can probe the exact imaginary-time dynamics with shallower circuits and large approximation error. Putting these scaling forms together, we can investigate universal properties of interesting quantum systems on quantum computers in a scalable and efficient way.

\begin{figure}[t]\centering
	\includegraphics[width=0.48\textwidth]{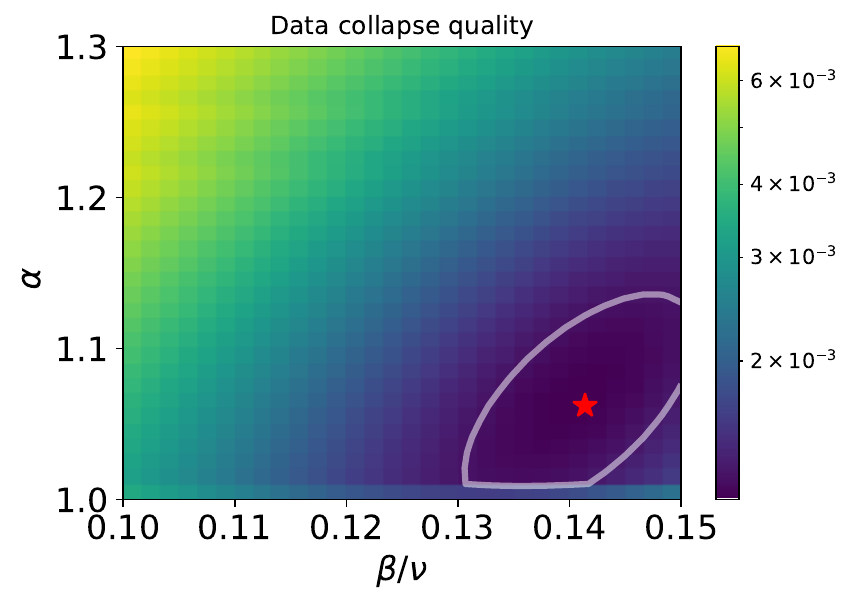}
	\caption{Data collapse quality grid search with variational imaginary time dynamics data of size $N=8,10,12$, circuit depth $D=2,3,4,5$ and time $\tau/N=0.1,0.2, 0.3, 0.4$. The optimal exponents estimate is $\alpha=1.06\pm0.03$ and $\beta/\nu=0.14\pm 0.05$. The grey contour indicates the boundary where the cost function is $10\%$ worse than the optimal estimate.}
\label{fig:sm-small-scaling}
\end{figure}

\end{widetext}

\end{document}